\documentclass[twocolumn, tighten]{aastex7}

\usepackage{amsmath}

\shorttitle{Ly$\alpha$ in Compact Broad-Line Emitters and LRDs}
\shortauthors{Y.\ Zhu et al.}

\begin{document}

\title{Ly$\alpha$ in Compact Broad-Line Emitters, Including Little Red Dots, Is More Closely Linked to the Host Galaxy than to the Broad-Line Nucleus}

\author[0000-0003-3307-7525]{Yongda Zhu}
\altaffiliation{These authors contributed equally to the analysis in this work.}
\affiliation{Steward Observatory, University of Arizona, 933 North Cherry Avenue, Tucson, AZ 85721, USA}
\email[show]{yongdaz@arizona.edu}

\author[0009-0003-5402-4809]{Zheng Ma}
\altaffiliation{These authors contributed equally to the analysis in this work.}
\affiliation{Steward Observatory, University of Arizona, 933 North Cherry Avenue, Tucson, AZ 85721, USA}
\email{mazh@arizona.edu}

\author[0000-0001-6052-4234]{Xiaojing Lin}
\affiliation{Department of Astronomy, Tsinghua University, Beijing 100084, China}
\email{linxj21@mails.tsinghua.edu.cn}

\author[0000-0002-1574-2045]{Junyu Zhang}
\affiliation{Steward Observatory, University of Arizona, 933 North Cherry Avenue, Tucson, AZ 85721, USA}
\email{junyuzhang@arizona.edu}

\author[0000-0002-6221-1829]{Jianwei Lyu}
\affiliation{Steward Observatory, University of Arizona, 933 North Cherry Avenue, Tucson, AZ 85721, USA}
\email{jianwei@arizona.edu}

\author[0000-0003-3310-0131]{Xiaohui Fan}
\affiliation{Steward Observatory, University of Arizona, 933 North Cherry Avenue, Tucson, AZ 85721, USA}
\email[]{xfan@arizona.edu}

\author[0000-0003-1344-9475]{Eiichi Egami}
\affiliation{Steward Observatory, University of Arizona, 933 North Cherry Avenue, Tucson, AZ 85721, USA}
\email{egami@arizona.edu}

\author[0000-0002-8651-9879]{Andrew J.\ Bunker}
\affiliation{Department of Physics, University of Oxford, Denys Wilkinson Building, Keble Road, Oxford OX1 3RH, UK}
\email{andy.bunker@physics.ox.ac.uk}

\author[0009-0003-4133-0292]{Sijia Cai}
\affiliation{Department of Astronomy, Tsinghua University, Beijing 100084, China}
\email{caisj23@mails.tsinghua.edu.cn}

\author[0000-0001-5211-1958]{Laura C. Keating}
\affiliation{Institute for Astronomy, University of Edinburgh, Blackford Hill, Edinburgh, EH9 3HJ, UK}
\email[]{Laura.Keating@ed.ac.uk}

\author[0000-0002-7579-2829]{Joseph Shields}
\affiliation{Large Binocular Telescope Observatory, Tucson, AZ, USA}
\email[]{}

\author[0000-0001-7673-2257]{Zhiyuan Ji}
\affiliation{Steward Observatory, University of Arizona, 933 North Cherry Avenue, Tucson, AZ 85721, USA}
\email{zhiyuanji@arizona.edu}

\author[0000-0002-5768-738X]{Xiangyu Jin}
\affiliation{Department of Astronomy, University of Michigan, 1085 S. University Ave., Ann Arbor, MI 48109, USA}
\email{jxiangyu@umich.edu}

\author[0000-0003-3762-7344]{Weizhe Liu}
\affiliation{Steward Observatory, University of Arizona, 933 North Cherry Avenue, Tucson, AZ 85721, USA}
\email[]{}

\author[0009-0006-5305-8363]{Suprabhas Narisetty}
\affiliation{Department of Astronomy, University of Illinois Urbana-Champaign, 1002 W. Green Street, Urbana, IL 61801, USA}
\email[]{ssn13@illinois.edu}

\author[0000-0003-2303-6519]{George H.\ Rieke}
\affiliation{Steward Observatory, University of Arizona, 933 North Cherry Avenue, Tucson, AZ 85721, USA}
\email{grieke@arizona.edu}

\author[0000-0002-7893-6170]{Marcia J.\ Rieke}
\affiliation{Steward Observatory, University of Arizona, 933 North Cherry Avenue, Tucson, AZ 85721, USA}
\email{mrieke@arizona.edu}

\author[0000-0002-5104-8245]{Pierluigi Rinaldi}
\affiliation{Space Telescope Science Institute, 3700 San Martin Drive, Baltimore, Maryland 21218, USA}
\email{prinaldi@stsci.edu}

\author[0000-0002-4622-6617]{Fengwu Sun}
\affiliation{Department of Astronomy, Westlake University, Hangzhou 310030, Zhejiang Province, China}
\affiliation{Center for Astrophysics $|$ Harvard \& Smithsonian, 60 Garden St., Cambridge, MA 02138, USA}
\email{sunfengwu@westlake.edu.cn}

\begin{abstract}
JWST has found many faint broad-line AGN at $z>4$, but it is unclear whether the Ly$\alpha$ we see from them comes from the accreting black hole or from its host galaxy. The answer matters for how these sources are interpreted and for how many ionizing photons they release. We use deep JWST/NIRSpec spectroscopy from the DIVER program to measure Ly$\alpha$ in 12 compact broad-line emitters (CBEs) at $z=4.2$--5.5 in GOODS-N. CBEs include little red dots (LRDs), little blue dots, and other broad-line AGN. Six are detected in Ly$\alpha$. Every source shows a broad-line region, yet Ly$\alpha$ does not follow it: the sources with the strongest H$\alpha$ relative to the UV include weak detections and nondetections, and every detected Ly$\alpha$ line (FWHM $\approx300$--1000~km~s$^{-1}$) is narrower than broad H$\alpha$ (1100--3300~km~s$^{-1}$). Instead, Ly$\alpha$ appears to track the host galaxy. We model the AGN-subtracted host with a dust-attenuated SED fit. Including the upper limits, the observed Ly$\alpha$ follows the Ly$\alpha$ production in this host model more closely than it follows H$\alpha$/UV, in both equivalent width and luminosity. Every detection exceeds its host-model value by a factor of 1.5--18, leaving room for a nuclear contribution. The black hole is visible in these sources, but the Ly$\alpha$ we observe is more closely tied to its host. The median effective Ly$\alpha$ escape fraction of the detections is only 0.09, which suggests that ionizing escape from CBEs is typically modest and varies widely.
\end{abstract}


\section{Introduction}\label{sec:intro}

JWST has revealed a large population of faint broad-line active galactic nuclei (AGN) at $z>4$ \citep[e.g.,][]{matthee_little_2024,juodzbalis_jades_2026,zhang_abundant_2026,lin_luminosity_2025,lin_bridging_2026}. Most are identified through broad H$\alpha$ emission from dense gas moving at thousands of km~s$^{-1}$ near an accreting black hole. Many are also X-ray weak and show little variability, which keeps their physical nature under debate \citep{yue_stacking_2024,maiolino_jwst_2024,kokubo_challenging_2025,greene_what_2025}. A prominent subset, the little red dots (LRDs), pairs compact red rest-optical continua with blue rest-UV emission. LRDs have been studied through their identification, sizes, number densities, and dust emission \citep{baggen_small_2024,perez-gonzalez_what_2024,williams_galaxies_2024,ma_no_2025,delvecchio_active_2025,ding_learning_2026}, black-hole and host masses \citep{juodzbalis_direct_2025,jones_m_rm_2025,ivey_cliff_2026}, rest-UV and rest-optical spectra \citep{akins_strong_2024,rinaldi_deciphering_2025,davis_oceans_2026,wang_water_2026}, low-redshift analogs \citep{burke_too_2025,rinaldi_beyond_2025,lin_discovery_2026,lin_lrds2_2026}, and possible origins and descendants \citep{jeon_little_2025,obuchi_discovery_2026,zhong_blackbody_2026}. If these AGN are common and their ionizing photons escape efficiently, they could contribute substantially to the ionizing background during reionization. Whether their Ly$\alpha$ and Lyman-continuum (LyC) photons escape depends on gas beyond the visible broad-line region (BLR).

Ly$\alpha$ traces high-redshift galaxies and the ionization state of the intergalactic medium \citep[IGM; e.g.,][]{mason_universe_2018,chen_impact_2026,tang_jwstnirspec_2024,christenson_relationship_2023,hashemi_ly_2025}. Ly$\alpha$ and LyC photons may escape through the same low-column-density, low-attenuation channels, and observations and simulations find a broad connection between their escape fractions \citep{lin_quantifying_2024,begley_connecting_2024,choustikov_great_2024}. The observed Ly$\alpha$ also depends on transfer through neutral gas, dust, the circumgalactic medium (CGM), and the IGM \citep[e.g.,][]{bolan_inferring_2022,chen_spurs_2026}, so it is a useful but indirect probe of LyC escape \citep[for reviews, see][]{ouchi_observations_2020,hayes_sizes_2023}. In the Case B limit, the intrinsic Ly$\alpha$/H$\alpha$ ratio is fixed to first order \citep{osterbrock_astrophysics_2006}, although optically thick or density-bounded gas can move it away from Case B \citep{scarlata_universal_2024,mcclymont_density-bounded_2025}. Because JWST measures both H$\alpha$ and Ly$\alpha$ in AGN at $z>4$, broad-line AGN offer a direct test of whether strong nuclear recombination emission comes with strong observed Ly$\alpha$. A type-1 view shows the BLR along one sightline, but Ly$\alpha$ photons scatter through gas from nuclear to intergalactic scales and need not escape along the same path.

Recent JWST observations show that the two can differ. LRDs and other broad-line sources can have strong broad Balmer emission but weak, narrow, or spatially extended Ly$\alpha$ \citep{torralba_weak_2026,ji_through_2026}, and broad Ly$\alpha$ is uncommon even when broad H$\alpha$ is clearly present \citep{kageura_atlas_2026}. Some LRDs may still have a BLR contribution: a stacked medium-resolution spectrum shows a broad Ly$\alpha$ wing that matches broad H$\alpha$ \citep{geris_little_2026}. Compact blue broad-line sources, or little blue dots \citep[LBDs;][]{madau_little_2026}, show higher Ly$\alpha$ visibility, possibly through lower gas columns \citep{mascia_ionizing_2026}. LRD selections still differ between studies, which use different color cuts or continuum-slope criteria \citep{greene_uncover_2024,kokorev_census_2024,kocevski_rise_2025,akins_cosmos-web_2025,hainline_investigation_2025,zhang_abundant_2026}, and spectroscopic LRD samples are heterogeneous \citep{de_graaff_little_2025,perez-gonzalez_little_2026}. To avoid depending on a particular LRD definition, we refer to LRDs, LBDs, and other broad-line AGN collectively as compact broad-line emitters (CBEs). This umbrella term spans the full range of continuum colors and is broader than the blue CBE selection of \citet{mascia_ionizing_2026}; we select our sample on broad H$\alpha$ alone.

The largest study so far analyzed 110 LRDs with NIRSpec PRISM spectra \citep{ji_through_2026}. It finds that Ly$\alpha$/H$\alpha$ lies below that of star-forming galaxies and that Ly$\alpha$ luminosity tracks [\ion{O}{3}] luminosity, which suggests that Ly$\alpha$ is tied to the host rather than to the compact broad-line source. Two questions remain open. At PRISM resolution, the Ly$\alpha$ line width cannot be compared with the broad-H$\alpha$ width, and [\ion{O}{3}] can be excited by both star formation and the AGN \citep{sok_constraints_2026,geris_little_2026}, so it may not fully isolate the host.

In this paper, we ask which component of a CBE is most closely related to the Ly$\alpha$ that we observe. Broad H$\alpha$ traces the visible broad-line nucleus. AGN-subtracted imaging and spectral energy distribution (SED) fitting give an independent view of the host galaxy, which in LRDs is often faint, extended, or offset from the compact source \citep{chen_host_2025,chen_physical_2025,rinaldi_not_2025}. We use the Deep Insights into UV Spectroscopy at the Epoch of Reionization (DIVER) survey to measure Ly$\alpha$ in 12 CBEs at $z=4.17$--5.54, mostly from medium-resolution G140M spectra that resolve line widths well below those of broad H$\alpha$. We test whether Ly$\alpha$ follows H$\alpha$, whether it shares the broad-H$\alpha$ line width, and whether it follows the Ly$\alpha$ in the AGN-subtracted host SED model. Section~\ref{sec:data} describes the data and measurements, Section~\ref{sec:results} presents the results, Section~\ref{sec:discussion} discusses the origin of the observed Ly$\alpha$ and the implications for ionizing escape, and Section~\ref{sec:summary} summarizes our findings. We adopt the Planck 2018 flat $\Lambda$CDM cosmology as implemented in Astropy \citep[$H_0=67.66$~km~s$^{-1}$~Mpc$^{-1}$, $\Omega_{\rm m}=0.310$;][]{planck_collaboration_planck_2020}. All magnitudes are in the AB system.

\begin{deluxetable*}{lcccccccc}[!ht]
\tablecaption{CBE sample and Ly$\alpha$ constraints \label{tab:sample}}
\tabletypesize{\scriptsize}
\tablewidth{0pt}
\tablehead{
\colhead{ID} &
\colhead{$z$} &
\colhead{Spectrum} &
\colhead{$F_{\rm Ly\alpha}$} &
\colhead{EW$_{\rm Ly\alpha}$} &
\colhead{$\log f_{\rm esc,Ly\alpha}^{\rm eff}$} &
\colhead{$\log L_{\rm H\alpha,total}$} &
\colhead{FWHM$_{\rm H\alpha,broad}$} &
\colhead{FWHM$_{\rm Ly\alpha,obs}$}
}
\startdata
1029154 & 4.17 & PRISM & $<2.88$ & $<59$ & $<-1.80$ & 42.57 & 2003 & \nodata \\
1086784 & 4.405 & G140M-2 & $<1.47$ & $<30$ & $<-1.84$ & 42.37 & 3179 & \nodata \\
1008411 & 4.41 & G140M-2 & $8.43\pm1.61$ & $273\pm71$ & $-0.99$ & 42.28 & 3281 & 424 \\
1008671 & 4.41 & G140M-2 & $<10.17$ & $<127$ & $<-1.28$ & 42.65 & 2272 & \nodata \\
1086855 & 4.41 & G140M-2 & $11.52\pm0.64$ & $360\pm72$ & $-1.11$ & 42.53 & 1724 & 961 \\
1033320 & 4.48 & PRISM & $<8.86$ & $<116$ & $<-0.91$ & 42.24 & 1951 & \nodata \\
1020621 & 4.681 & G140M-2 & $10.25\pm0.40$ & $216\pm57$ & $-0.88$ & 42.32 & 1605 & 316 \\
1034620 & 5.19 & G140M-1 & $11.56\pm0.42$ & $73\pm4$ & $-1.20$ & 42.80 & 1077 & 512 \\
1077652 & 5.229 & G140M-1 & $<1.10$ & $<20$ & $<-1.68$ & 42.26 & 1026 & \nodata \\
9994014 & 5.23 & G140M-1 & $<0.70$ & $<10$ & $<-2.38$ & 42.77 & 2084 & \nodata \\
1020514 & 5.36 & G140M-1 & $161.9\pm2.6$ & $693\pm19$ & $0.08$ & 42.70 & 1612 & 585 \\
1087388 & 5.54 & G140M-1 & $1.79\pm0.25$ & $24\pm4$ & $-2.80$ & 43.65 & 2964 & 665 \\
\enddata
\tablecomments{Ly$\alpha$ fluxes are in $10^{-18}$~erg~s$^{-1}$~cm$^{-2}$, EWs are rest-frame \AA, FWHMs are in km~s$^{-1}$, and luminosities are in erg~s$^{-1}$. Limits are 2$\sigma$. G140M-1 and G140M-2 denote first- and second-order G140M; PRISM is first order. FWHM$_{\rm Ly\alpha,obs}$ includes instrumental broadening (Section~\ref{sec:results_profiles}). $f_{\rm esc,Ly\alpha}^{\rm eff}$ uses total H$\alpha$. Broad-H$\alpha$ FWHMs are from \citet{zhang_abundant_2026} and \citet{juodzbalis_jades_2026}.}
\end{deluxetable*}

\begin{figure*}[!ht]
    \centering
    \includegraphics[width=0.94\textwidth]{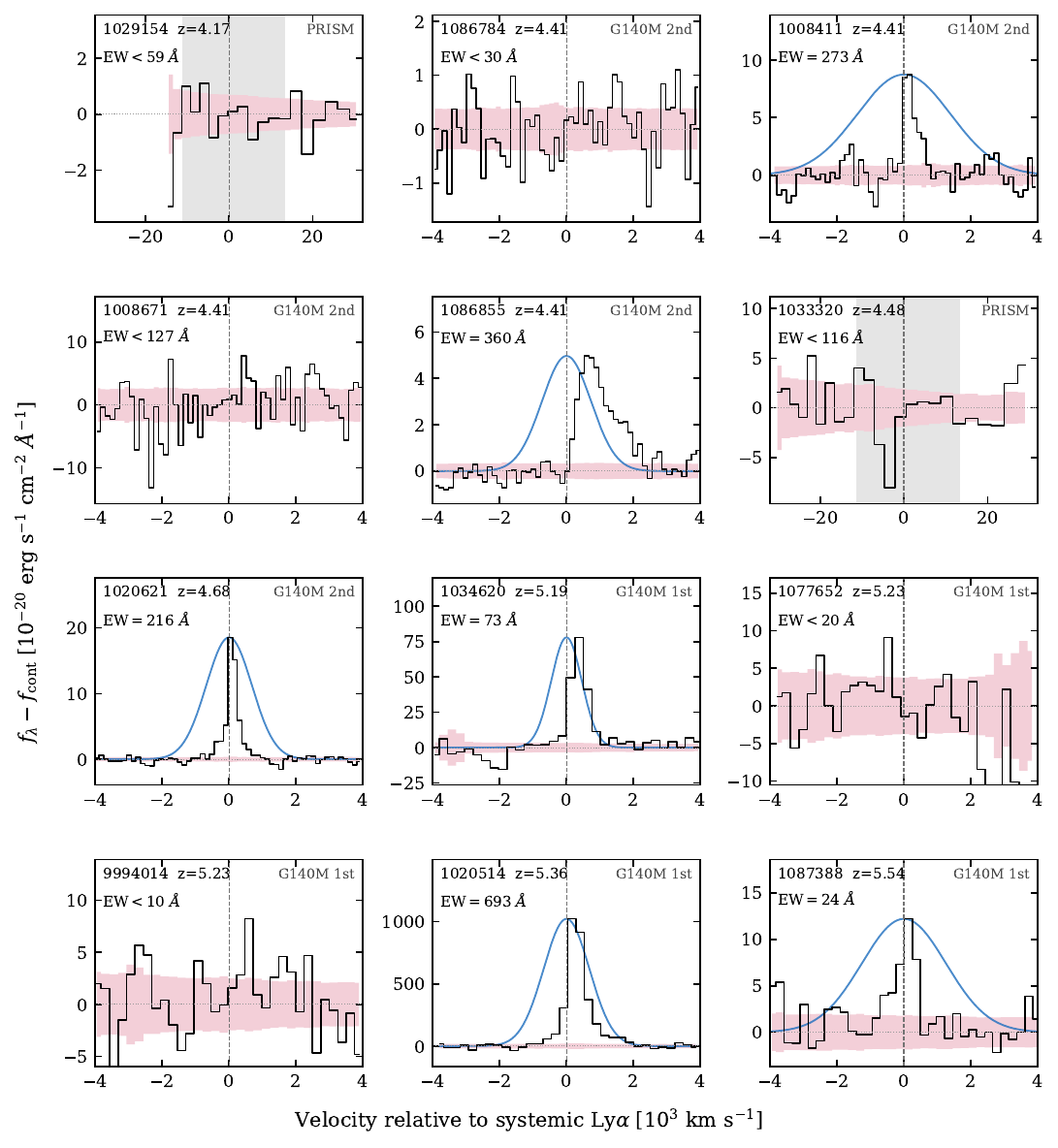}
    \caption{
    Ly$\alpha$ spectra of the 12 CBEs, ordered by redshift, in velocity relative to the systemic (H$\alpha$) redshift. Labels give the adopted spectrum and the rest-frame EW$_{\rm Ly\alpha}$ (2$\sigma$ limits for nondetections). Pink bands show the $1\sigma$ uncertainty, and the local continuum is subtracted. G140M panels span $\pm4000$~km~s$^{-1}$; second-order panels show the apparent, first-order-calibrated flux density, while the adopted line fluxes include the order-dependent response conversion. For G140M detections, blue curves show a Gaussian with the broad-H$\alpha$ FWHM at the systemic velocity, scaled to the Ly$\alpha$ peak. The two PRISM panels span $\pm32{,}000$~km~s$^{-1}$, and the gray band marks the rest-frame 1170--1270~\AA\ integration window. For 1033320, second-order G140M Ly$\alpha$ falls in a spectral gap, and we adopt the PRISM constraint.
    }
    \label{fig:overview}
\end{figure*}

\section{Data and Measurements}\label{sec:data}

\subsection{DIVER spectroscopy}\label{sec:data_spec}

DIVER (JWST PID 8018; PI: X. Lin) is a NIRSpec multi-object spectroscopic survey of GOODS-N, with median exposure times of 20.4~hr in F070LP/G140M and 2.3~hr in PRISM. Target selection and slit allocation did not use prior information on Ly$\alpha$. The reduction follows \citet{zhu_smiles_2026}, with the DIVER-specific updates of \citet{zhu_low_2026}. These updates extend the F070LP/G140M wavelength coverage to $\simeq3~\mu$m, using the flat-field and sensitivity calibration of \citet{scholtz_jades_2025}. With this calibration, fluxes agree with those from the other gratings within 10\%. The pipeline treats the nodded MSA spectra as point sources and applies the standard point-source path-loss correction for the planned intra-shutter position. We extract 1D spectra in a six-pixel ($\approx0.6\arcsec$) aperture and apply no further aperture correction or scaling to imaging. Ly$\alpha$ is likely more extended than the UV continuum. Emission scattered outside the $0.2\arcsec$ shutter or removed by the nodded background subtraction is therefore potentially missed, and line fluxes and EWs might be biased low. \citet{napolitano_ly_2026} found an average NIRSpec Ly$\alpha$ slit loss of $35\pm10\%$ at $z\simeq6$. All sources are reduced and measured in the same way, so this bias affects relative comparisons less than absolute values.

F070LP/G140M covers first-order Ly$\alpha$ at $z\gtrsim4.8$. At lower redshift, Ly$\alpha$ is recorded in second order at an apparent wavelength $\lambda_{\rm app}=2\times1215.67\,(1+z)$~\AA. This falls in the extended range, whose calibration assumes first-order light. We convert second-order line fluxes to true fluxes with the ratio of the first- and second-order sensitivity curves of msaexp \citep{brammer_msaexp_2023} and the wavelength Jacobian, $d\lambda_{\rm app}=2\,d\lambda_{\rm true}$, for apparent wavelengths of 1.30--2.00~$\mu$m.
The conversion factors for our five second-order measurements are 3.19--5.05, and a $\pm0.002~\mu$m shift in apparent wavelength changes them by $\lesssim4\%$. We test the conversion with sources at $z>4.8$, whose Ly$\alpha$ appears in both orders of the same G140M spectrum. For the 13 with second-order ${\rm S/N}\ge5$, at true wavelengths of 0.75--0.98~$\mu$m, the converted second-order fluxes are a median of 1.01 times the first-order fluxes (16th--84th percentiles 0.69--1.94). Our second-order lines lie at 0.66--0.69~$\mu$m, where the second-order sensitivity rises steeply, so individual fluxes may be uncertain by a factor of $\sim2$. Halving or doubling them does not change our conclusions (Sections~\ref{sec:results_host} and \ref{sec:discussion_ionizing}). PRISM gives only marginal Ly$\alpha$ fluxes for 1086855 and 1008411 (S/N of 2.8 and 1.1), 0.5 and 0.2 times the converted values. Near the blue end of PRISM, Ly$\alpha$ fluxes of other DIVER sources scatter widely and include significantly negative values, so we do not use PRISM to test the conversion.

\begin{figure*}[!ht]
    \centering
    \includegraphics[width=\textwidth]{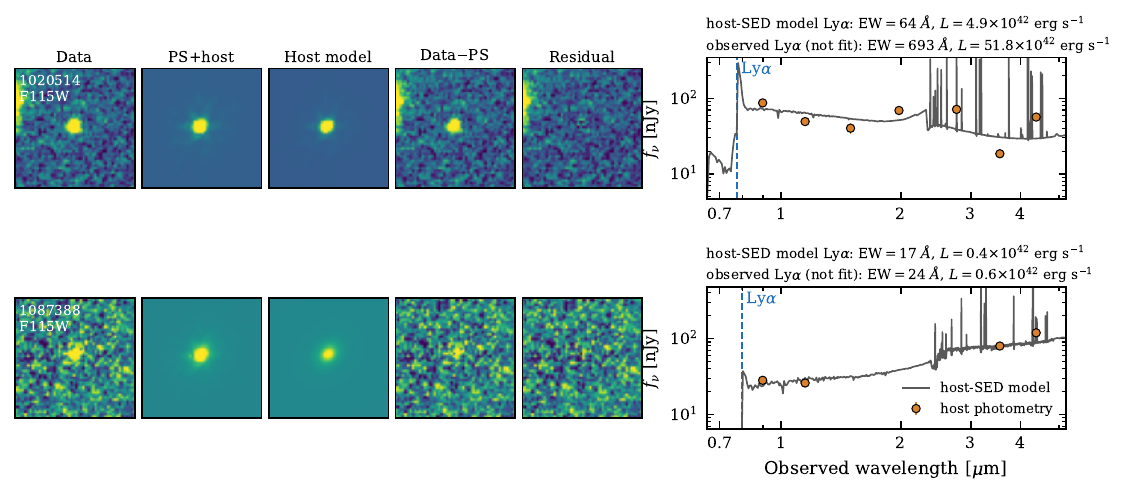}
    \caption{
    AGN--host decomposition and host SED for a strong (1020514, top) and a weak (1087388, bottom) Ly$\alpha$ detection. Left: F115W data, the PS+host model, the S\'ersic host model, the data after PS subtraction, and the residual \citep{ma_undermassive_2026}; each panel has its own display stretch. Right: AGN-subtracted host photometry (orange) and the best-fit host SED model (gray), with the Ly$\alpha$ wavelength marked; 1087388 has no JADES F150W, F200W, or F277W imaging, so its host photometry has four bands. The annotations compare the Ly$\alpha$ in the host SED model with the observed Ly$\alpha$, which is not used in the fit.
    }
    \label{fig:decomp}
\end{figure*}

\subsection{Sample}\label{sec:data_sample}

The sample contains the 12 sources with secure broad H$\alpha$ in the CONGRESS/FRESCO census of \citet{zhang_abundant_2026} and the JADES census of \citet{juodzbalis_jades_2026} that have usable DIVER Ly$\alpha$ coverage (Table~\ref{tab:sample}). We restrict the sample to $z<6$, because IGM attenuation of Ly$\alpha$ rises steeply at higher redshift \citep[e.g.,][]{jones_jades_2025,zhu_low_2026,chen_impact_2026}. Our test requires a directly visible BLR, so the sample excludes photometric LRD candidates without broad-line spectroscopy and narrow-line AGN. \citet{zhang_abundant_2026} identify the LRDs among these broad-H$\alpha$ emitters with color selections \citep[see also][]{kocevski_rise_2025} and find that 42\%--47\% of them are not LRDs. We analyze the CBEs as one population and do not classify individual objects.

\subsection{\texorpdfstring{Ly$\alpha$ measurements}{Lyalpha measurements}}\label{sec:data_lya}

Figure~\ref{fig:overview} shows the Ly$\alpha$ spectra of the 12 CBEs. We measure first-order G140M Ly$\alpha$ following \citet{zhu_low_2026}. A local continuum is fitted outside the line, and the line peak is located within $\pm1000$~km~s$^{-1}$ of the systemic redshift. The continuum-subtracted flux is summed within $\pm16$~\AA\ (observed frame; $\approx\pm600$~km~s$^{-1}$) of the peak. A line is detected if both the integrated and peak signal-to-noise ratios are $\ge3$. Second-order fluxes are summed within $\pm2500$~km~s$^{-1}$ after subtracting the median continuum at $+4000$ to $+9000$~km~s$^{-1}$, and then converted. For PRISM, we subtract a clipped red-side continuum at rest-frame 1350--1500~\AA\ and integrate over 1170--1270~\AA. Nondetections are reported as 2$\sigma$ upper limits. We adopt PRISM for 1029154, whose second-order Ly$\alpha$ falls at 1.257~$\mu$m, below the calibrated range, and for 1033320, whose second-order Ly$\alpha$ falls in a spectral gap.

The rest-frame EW$_{\rm Ly\alpha}$ divides the line flux by the continuum at rest-frame 1300~\AA\ from the best-fit SED of the total (AGN+host) photometry \citep{zhu_low_2026}; the continuum is largely undetected in the grating spectra. For 1034620, which lacks an SED fit, we use the PRISM red-side continuum.

We define the effective Ly$\alpha$ escape fraction
\begin{equation}
f_{\rm esc,Ly\alpha}^{\rm eff}=\frac{L_{\rm Ly\alpha}}{8.7\,L_{\rm H\alpha}},
\end{equation}
where $L_{\rm H\alpha}$ is the total (broad+narrow) H$\alpha$ luminosity, not corrected for dust, and 8.7 is the Case B Ly$\alpha$/H$\alpha$ ratio \citep{osterbrock_astrophysics_2006} for $T_e=10^4$~K and $n_e\approx350$~cm$^{-3}$ \citep{hayes_lyman_2015}. $f_{\rm esc,Ly\alpha}^{\rm eff}$ is an empirical visibility ratio that includes radiative transfer in the nucleus and host, CGM and IGM transmission, aperture effects, and the H$\alpha$ normalization. Total H$\alpha$ is from the NIRCam grism measurements of \citet{zhang_abundant_2026}. For the JADES sources 1020621 and 1077652, we fit narrow and broad H$\alpha$ and [N\,{\sc ii}] in the public JADES G395M spectra \citep{deugenio_jades_2025}; the narrow component carries about two-thirds of the H$\alpha$ flux of 1020621. For 1020514, $f_{\rm esc,Ly\alpha}^{\rm eff}=1.20$; the aperture mismatch between the slit Ly$\alpha$ and slitless H$\alpha$ measurements and departures from Case B, such as collisional enhancement of Ly$\alpha$, may contribute.

\begin{figure}[!t]
\centering
\includegraphics[width=\columnwidth]{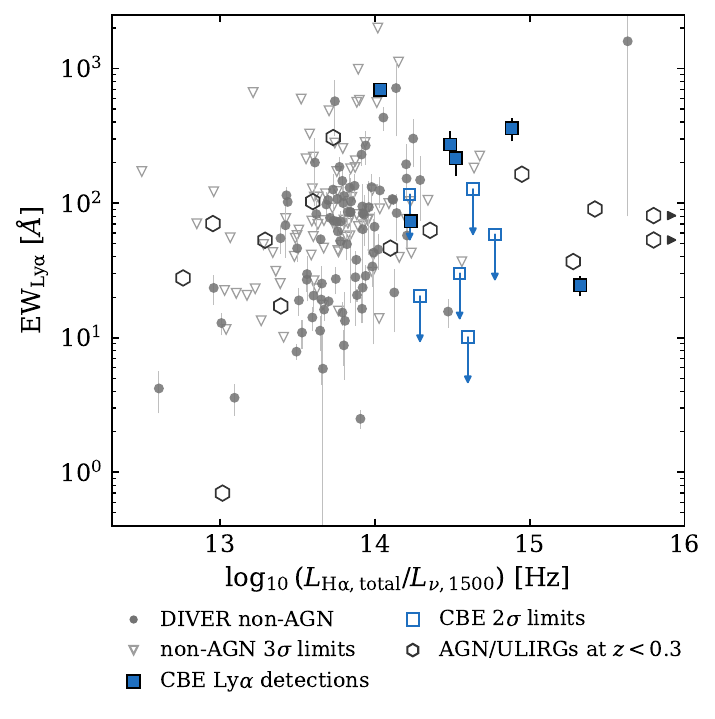}
\caption{
Ly$\alpha$ EW versus H$\alpha$/UV. Both samples use total H$\alpha$ and the observed total $L_{\nu,1500}$. Gray circles and open triangles show DIVER non-AGN detections and 3$\sigma$ upper limits. Blue filled squares show CBE Ly$\alpha$ detections, and open squares with arrows show 2$\sigma$ upper limits. Error bars show 1$\sigma$ EW uncertainties. Open hexagons show Ly$\alpha$-detected AGN and ULIRGs at $z<0.3$ from \citet{liu_galactic_2022}, whose UV continuum is measured near 1125~\AA, so their positions are approximate. Two of them, at $\log(L_{\rm H\alpha}/L_\nu)=16.2$ and 16.7, are shown at the right edge with arrows.
}
\label{fig:halpha_uv}
\end{figure}

\subsection{\texorpdfstring{Host decomposition and host-model Ly$\alpha$}{Host decomposition and host-model Lyalpha}}\label{sec:data_host}

We fit the AGN--host decomposition of all 12 sources in the JADES Data Release 5 (DR5) NIRCam images and catalogs \citep{eisenstein_overview_2023,johnson_jwst_2026,robertson_jwst_2026}, generally following \citet{ma_undermassive_2026} (Figure~\ref{fig:decomp}). The AGN is modeled with the JADES mosaic point-spread function (PSF) and the host with a single PSF-convolved S\'ersic profile. To improve the host characterization, we fit the short-wavelength bands jointly, with one S\'ersic shape and one AGN--host offset shared across them. A Markov chain Monte Carlo fit then gives the point-source (PS) and host fluxes in each band. We fit the host photometry with Prospector \citep{johnson_stellar_2021} and FSPS \citep{conroy_propagation_2009}, using the model of \citet[][Section 2.4]{zhu_low_2026}. The model assumes a \citet{chabrier_galactic_2003} initial mass function and a seven-bin nonparametric star-formation history with the continuity prior of \citet{leja_how_2019}. It includes the flexible dust law of \citet{kriek_dust_2013} with extra attenuation of young stars \citep{charlot_simple_2000}, nebular emission \citep{byler_nebular_2017}, dust emission \citep{draine_infrared_2007}, and a mid-infrared AGN component. The redshift may vary within $\pm0.01$ of the spectroscopic value, and the model photometry is computed with the neural-network emulator of \citet{mathews_as_2023}. The fitted photometry begins at F090W, redward of Ly$\alpha$ for all sources except 1087388, whose Ly$\alpha$ red wing falls at the blue edge of F090W and contributes $\lesssim4\%$ of that band. Neither the spectroscopic Ly$\alpha$ nor H$\alpha$ enters the fit.

The host-model Ly$\alpha$ is the Ly$\alpha$ emission in the dust-attenuated best-fit host SED model. We integrate the positive excess above a power-law continuum, fitted at rest-frame 1275--1305 and 1325--1365~\AA, over 1215--1270~\AA, and convert it to an EW and a luminosity. The model includes dust attenuation but no Ly$\alpha$ resonant transfer through neutral gas, the CGM, or the IGM, so the host-model Ly$\alpha$ traces Ly$\alpha$ production in the host. The model is sampled at $\approx10$~\AA\ near Ly$\alpha$, so the integration is approximate. The host model of 1086784 has no coherent Ly$\alpha$ excess, and we omit it from the host comparison.

\begin{figure*}[!t]
\centering
\includegraphics[width=\textwidth]{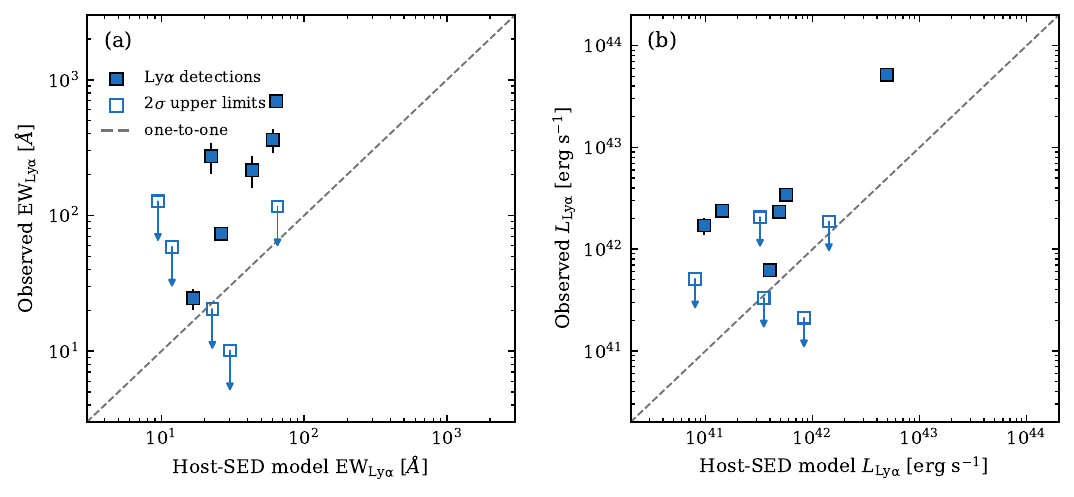}
\caption{
Observed Ly$\alpha$ versus the Ly$\alpha$ emission in the dust-attenuated best-fit host SED model, in (a) EW and (b) luminosity. The host model uses neither the spectroscopic Ly$\alpha$ nor photometry containing Ly$\alpha$. Filled squares show the six detections with 1$\sigma$ error bars, and open squares with arrows show 2$\sigma$ upper limits. The dashed line marks equality. 1086784 has no coherent host-model Ly$\alpha$ and is omitted.
}
\label{fig:host_lya}
\end{figure*}

\subsection{Non-AGN comparison sample}\label{sec:data_controls}

The comparison sample is the 250 non-AGN galaxies in the DIVER Ly$\alpha$ sample of \citet{zhu_low_2026}, measured with the same spectra and methods. None of the CBEs are in this sample. We use the 173 galaxies with H$\alpha$ and UV measurements: 84 Ly$\alpha$ detections and 89 3$\sigma$ upper limits, as in \citet{zhu_low_2026}. At the 2$\sigma$ level used for the CBEs, these limits would be a factor of 1.5 lower.

\section{Results}\label{sec:results}

The 12 sources include six Ly$\alpha$ detections and six 2$\sigma$ upper limits, a detection fraction of 0.50 (68\% binomial confidence interval 0.36--0.64). Non-AGN galaxies at the same redshifts have a similar detection fraction of 0.40 (34 of the 86 with Ly$\alpha$ measurements at $z=4.17$--5.54; 0.34--0.45). The detections span EW$_{\rm Ly\alpha}=24$--693~\AA, and the limits span EW$_{\rm Ly\alpha}<10$ to $<127$~\AA.

\subsection{\texorpdfstring{Strong H$\alpha$ does not predict strong Ly$\alpha$}{Strong Halpha does not predict strong Lyalpha}}\label{sec:results_halpha}

Figure~\ref{fig:halpha_uv} shows EW$_{\rm Ly\alpha}$ as a function of $L_{\rm H\alpha}/L_{\nu,1500}$, where both samples use total H$\alpha$ and the observed total UV continuum. The CBEs have higher H$\alpha$/UV than most DIVER non-AGN galaxies, but larger H$\alpha$/UV does not imply stronger Ly$\alpha$. Source 1087388 has the largest H$\alpha$ luminosity and H$\alpha$/UV in the sample but EW$_{\rm Ly\alpha}=24$~\AA, while 1020514 has the strongest Ly$\alpha$ at the lowest H$\alpha$/UV among the CBEs.

Both axes are normalized by the UV continuum, so some positive covariance is expected. A fit to the 84 non-AGN detections gives
\begin{equation}
\log {\rm EW}_{\rm Ly\alpha}=a+b\log(L_{\rm H\alpha}/L_{\nu,1500}),
\end{equation}
with $a=-8.719$ and $b=0.757$. The six CBE detections scatter around this relation, with a median residual of $+0.03$~dex and individual residuals from $-1.49$ to $+0.94$~dex. Five of the six upper limits lie 0.25--1.33~dex below the relation, and the PRISM limit for 1033320 lies on it. At high H$\alpha$/UV, the CBEs span a wide range of Ly$\alpha$ strength, and several lie well below the non-AGN relation. The 14 Ly$\alpha$-detected AGN and ULIRGs at $z<0.3$ from \citet{liu_galactic_2022} show the same behavior: nine lie below the relation (median residual $-0.32$~dex), and all five with $\log(L_{\rm H\alpha}/L_{\nu})>14.9$ lie 0.4--2.0~dex below it. We find no clear trend between EW$_{\rm Ly\alpha}$ and UV luminosity, total H$\alpha$ luminosity, or broad-H$\alpha$ FWHM among the six detections (Spearman rank coefficient $|\rho|\le0.49$, $p$-value $\ge0.36$).

\subsection{\texorpdfstring{Observed Ly$\alpha$ and host-model Ly$\alpha$}{Observed Lyalpha and host-model Lyalpha}}\label{sec:results_host}

Because the Ly$\alpha$ strength does not follow H$\alpha$/UV, we turn to the host galaxy. Figure~\ref{fig:host_lya} compares the observed and host-model Ly$\alpha$ for the 11 sources with a host-model value (six detections and five upper limits). The observed Ly$\alpha$ tends to rise with the host-model Ly$\alpha$, more clearly in EW than in luminosity, but not with H$\alpha$/UV.

To include the upper limits, we use the generalized Kendall rank correlation coefficient $\tau$ \citep{isobe_statistical_1986}, which runs from $-1$ to $+1$ like Spearman $\rho$ and uses only pairs of sources whose order is unambiguous; $p$-values come from random permutations. The host model gives $\tau=0.37$ ($p=0.14$) in EW and $\tau=0.26$ ($p=0.31$) in luminosity, whereas H$\alpha$/UV gives $\tau=-0.19$ ($p=0.44$). In bootstrap resamplings, the host relation has the larger $\tau$ in 88\% of cases in EW (against H$\alpha$/UV) and in 77\% in luminosity (against broad-H$\alpha$ luminosity, $\tau=0.04$). These fractions are 71\%--93\% when we drop 1034620 and 9994014, whose decompositions are affected by a foreground neighbor and image artifacts \citep{ma_undermassive_2026}, or halve or double the second-order fluxes. A linear fit that treats the upper limits as censored values gives a positive slope of observed on host-model EW (one-sided permutation $p=0.03$), whereas H$\alpha$/UV gives none ($p=0.67$). With the slope fixed at unity, the host model predicts the observed EW with a scatter of 0.66~dex, compared with 1.05~dex for H$\alpha$/UV; in luminosity, the scatter is 0.81~dex for the host model and 1.07~dex for Case B H$\alpha$. With 11 sources, this preference is not yet decisive. The six detections alone give the same picture: Spearman $\rho=0.83$ ($p=0.058$) for the host model and $\rho=-0.49$ ($p=0.36$) for H$\alpha$/UV.

Every detection lies above the one-to-one relation, by factors of 1.5--18. Because the host model already includes dust attenuation and resonant transfer usually reduces Ly$\alpha$ further, the excess implies additional Ly$\alpha$ from outside the stellar host component, possibly from the AGN, or host Ly$\alpha$ production that the SED fit underestimates. The host models also underpredict narrow H$\alpha$. Their dust-attenuated H$\alpha$ is a median of 0.3 (range 0.06--1.15) of the observed narrow H$\alpha$. Part of the narrow-line emission may therefore arise outside the stellar host component, or the fit may miss some star formation. Conversely, 9994014 lies at least a factor of 4 below its host-model luminosity, so host Ly$\alpha$ production alone does not guarantee visible Ly$\alpha$.

Figure~\ref{fig:decomp} shows the two ends of the sample, 1020514 and 1087388. In both, the host accounts for most of the rest-frame 1500~\AA\ light (75\% and 64\%), and the point source dominates the rest-frame optical. Their host-model EWs are 64 and 17~\AA, and their observed EWs are 693 and 24~\AA. The host of 1020514 is young, low-mass, and nearly dust-free (mass-weighted age $\approx30$~Myr, $\log M_\ast/M_\odot=7.9$, and diffuse dust optical depth $\tau_V=0.01$), whereas the host of 1087388 is older, more massive, and dustier ($\approx400$~Myr, 9.2, and 0.36). Host age alone does not set the Ly$\alpha$ strength, however: among the detections, the oldest host, 1008411 ($\approx550$~Myr), has EW$_{\rm Ly\alpha}=273$~\AA.

\begin{figure}[!t]
\centering
\includegraphics[width=\columnwidth]{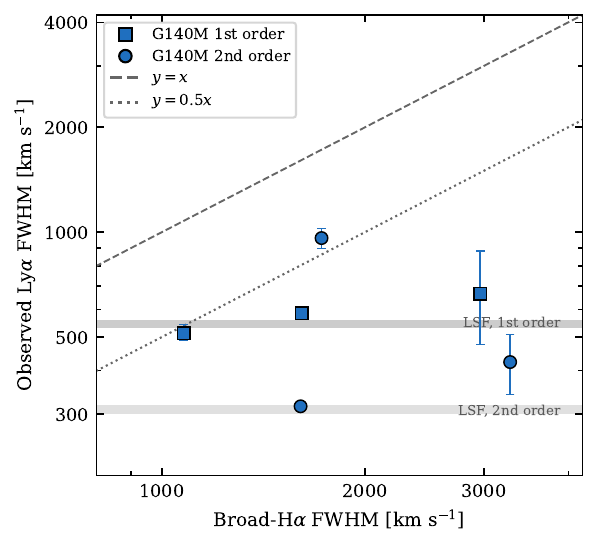}
\caption{
Observed Ly$\alpha$ FWHM versus broad-H$\alpha$ FWHM for the six detections, from first-order (squares) and second-order (circles) G140M spectra. The Ly$\alpha$ widths include instrumental broadening; gray bands mark the nominal instrumental FWHM for each order. The dashed line marks equal widths, and the dotted line marks $y=0.5x$, the approximate apparent width of a BLR-width line whose blue half is absorbed.
}
\label{fig:fwhm}
\end{figure}

\subsection{\texorpdfstring{Ly$\alpha$ is narrower than broad H$\alpha$}{Lyalpha is narrower than broad Halpha}}\label{sec:results_profiles}

If the observed Ly$\alpha$ came directly from the BLR, we would expect it to be about as broad as broad H$\alpha$. Figure~\ref{fig:fwhm} shows that it is not: every detected Ly$\alpha$ line is narrower than broad H$\alpha$, with observed FWHMs of 316--961~km~s$^{-1}$ against 1077--3281~km~s$^{-1}$, a ratio of 0.13--0.56. These Gaussian-fit widths include instrumental broadening (about 530--560~km~s$^{-1}$ in first order and 300--320~km~s$^{-1}$ in second order), so the intrinsic lines are narrower still; three lines (1020621, 1034620, and 1020514) are at most marginally resolved.

Two effects could make BLR Ly$\alpha$ look narrower. First, neutral gas can absorb the blue half of the line, roughly halving its apparent width (dotted line in Figure~\ref{fig:fwhm}). Only 1086855 lies above this line. Its Ly$\alpha$ is the broadest in the sample (${\rm FWHM}=961\pm64$~km~s$^{-1}$, peak redshifted by $\approx800$~km~s$^{-1}$) but still narrower than its broad H$\alpha$ (1724~km~s$^{-1}$). This line may reflect clumpy neutral gas or a favorable escape path, as inferred for an LRD with broad Ly$\alpha$ in SPURS \citep{tang_spurs_2026}. Second, a faint broad component could hide beneath a bright narrow core. When we add a systemic component with the broad-H$\alpha$ width, five lines do not need it ($2\sigma$ limits of 10--49\% of the narrow-line flux). Only 1020514, the brightest line, prefers a broad red wing with $\approx40\%$ of the narrow-line flux, which may be a weak nuclear contribution or a red scattering wing. \citet{geris_little_2026} detect a broad Ly$\alpha$ component in a stacked medium-resolution spectrum of eight LRDs; because a stack averages over sources, such a component could arise in a subset of objects like 1020514.

Ly$\alpha$ produced or scattered away from the nucleus could also be offset along the slit. In the 2D spectra, the Ly$\alpha$ and continuum positions agree within $0.14\arcsec$ ($\le1$~kpc) for all six detections; the formal $0.024\arcsec$ offset of 1020621, in second-order data, drops to $2.4\sigma$ when we allow for a 0.1-pixel systematic uncertainty.

\section{Discussion}\label{sec:discussion}

\subsection{\texorpdfstring{What sets the observed Ly$\alpha$}{What sets the observed Lyalpha}}\label{sec:discussion_origin}

Every source in our sample has a directly visible BLR, yet the Ly$\alpha$ strength does not follow the H$\alpha$ luminosity or H$\alpha$/UV. Other broad-line samples also show weak or mostly narrow Ly$\alpha$ \citep{torralba_weak_2026,kageura_atlas_2026}. In 110 LRDs at $z\ge4$, \citet{ji_through_2026} find Ly$\alpha$ luminosities and EWs similar to those of star-forming galaxies but systematically lower Ly$\alpha$/H$\alpha$. Our CBEs show this deficit mainly through their nondetections: the detections scatter around the non-AGN relation in Figure~\ref{fig:halpha_uv}, while five of the six upper limits lie below it. AGN and ULIRGs at $z<0.3$ also lie at or below this relation, which suggests that AGN do not generally have higher Ly$\alpha$ visibility than star-forming galaxies. Within that sample, Ly$\alpha$ EW increases with AGN luminosity and AGN fraction \citep{liu_galactic_2022}; we find no such trend with H$\alpha$ luminosity.

Two processes can separate the observed Ly$\alpha$ from the BLR. First, dense nuclear gas can change the intrinsic Ly$\alpha$/H$\alpha$ ratio: Balmer-continuum opacity, enhanced $n=2$ populations, collisional effects, and electron scattering can drive it far from Case B \citep{shields_implications_1993}. In LRDs, such dense gas may also produce the observed Balmer breaks and Balmer absorption \citep{inayoshi_extremely_2025,naidu_black_2025,de_graaff_remarkable_2025,de_graaff_little_2025}, and the broad Balmer profiles may partly reflect radiative transfer through it rather than BLR kinematics alone \citep{matthee_engine_2026}. Second, Ly$\alpha$ is resonantly transferred through the nucleus, host, and CGM. The neutral-gas covering fraction, dust, and velocity structure shape the emerging line \citep[e.g.,][]{gronke_resonant_2017}, and gas velocities move Ly$\alpha$ into regions of higher or lower IGM transmission \citep{cain_chasing_2025}. A high covering fraction can suppress Ly$\alpha$ even when the BLR is seen through a lower-opacity sightline. In LRDs, enhanced broad H$\alpha$ relative to bolometric luminosity suggests high BLR covering factors and large gas columns \citep{yanagisawa_scaling_2026}, which would favor this suppression. Where Ly$\alpha$ does escape, it can do so through narrow channels. In LRD-204851, radiative-transfer modeling of the Ly$\alpha$ profile favors a low-column-density biconical cavity in a dense, slowly expanding neutral envelope \citep{ji_compact_2026}, and narrow-line ratios in LRDs also point to lower-density channels through the gas envelope \citep{sok_constraints_2026}. The narrow line profiles (Section~\ref{sec:results_profiles}) suggest that the observed Ly$\alpha$ is largely shaped outside the BLR.

Observed Ly$\alpha$ follows the host-model Ly$\alpha$ more closely than H$\alpha$/UV (Section~\ref{sec:results_host}). Because the host model traces Ly$\alpha$ production in the host, this relation suggests that host star formation contributes to the observed line, although the host dust and gas that shape the host-model Ly$\alpha$ could also regulate the escape of nuclear Ly$\alpha$. \citet{ji_through_2026} reach a similar conclusion by a different route. In their LRDs, Ly$\alpha$ luminosity tracks [\ion{O}{3}] luminosity more closely than [\ion{O}{3}] EW, and at $z\gtrsim5.5$ the Ly$\alpha$ maps are often extended, asymmetric, and offset from the rest-optical light. They interpret these results as Ly$\alpha$ associated with the host-scale component rather than with the compact source of the broad lines and red continuum. In our 2D spectra, Ly$\alpha$ and continuum positions agree within $0.14\arcsec$ along the slit (Section~\ref{sec:results_profiles}), but the $0.2\arcsec$ shutter would miss emission extended or offset beyond it.

The excess above the host model leaves room for a nuclear contribution, which would also emerge narrow after transfer through host and CGM gas. We conclude that, in our sample, the observed Ly$\alpha$ of CBEs is more closely related to the host galaxy than to the BLR. Larger samples with deep rest-UV spectroscopy and Ly$\alpha$ maps will test this result beyond our six detections and five upper limits.

\subsection{Implications for ionizing escape}\label{sec:discussion_ionizing}

Ly$\alpha$ visibility gives a rough guide to ionizing escape, because LyC leakers at $z\simeq2$ show larger Ly$\alpha$ escape fractions, less dust, and lower neutral-gas columns \citep{naidu_synchrony_2022}. At $z\simeq5$--6, CGM and IGM transmission add scatter to this relation \citep{tang_ly_2024}. The six detections have a median $f_{\rm esc,Ly\alpha}^{\rm eff}=0.091$, a 16th--84th percentile range of 0.051--0.35, and a full range of 0.0016--1.20; the upper percentile is set mainly by 1020514. Five of the six upper limits lie below the median, and the PRISM limit for 1033320 ($<0.12$) lies above it. The detection-only median is an upper estimate of the characteristic escape scale; halving or doubling the second-order fluxes (Section~\ref{sec:data_spec}) gives 0.057--0.18. Normalizing by narrow H$\alpha$ alone, which carries 23--65\% of the total, raises the individual values by factors of 1.6--4.2 and the detection median to 0.28.

\citet{jiang_agns_2025} derive an upper bound on the AGN contribution to the ionizing budget at $z\approx7.5$ that scales with the assumed LyC escape fraction:
\begin{equation}
\frac{\dot{N}_{\rm ion}^{\rm AGN}}{\dot{N}_{\rm ion,req}}\simeq0.32\left(\frac{f_{\rm esc}^{\rm LyC}}{0.75}\right).
\end{equation}
Taking $f_{\rm esc,Ly\alpha}^{\rm eff}=0.091$ as an approximate escape scale gives a contribution of about 3.9\%, and the 16th--84th percentile range gives 2.2--15\%; the narrow-H$\alpha$ median gives about 12\%. This scaling is approximate: the normalization refers to $z\approx7.5$, our sample lies at $z\approx4$--5.5, and $f_{\rm esc,Ly\alpha}^{\rm eff}$ measures Ly$\alpha$ visibility. If ionizing photons also escape through narrow low-column channels like that in LRD-204851 \citep{ji_compact_2026}, the escape fraction depends on viewing angle, and our single sightline need not give the angle-averaged value. The star-forming-galaxy relation $f_{\rm esc}^{\rm LyC}\simeq0.15f_{\rm esc}^{\rm Ly\alpha}$ of \citet{begley_connecting_2024} would give an even smaller contribution of about 0.58\%. Estimates in which AGN supply a much larger share, such as 31--75\% at $z\simeq5$--6 \citep{singha_faint_2025}, depend on the adopted LyC escape fractions, which our data suggest are typically modest for CBEs. LRD ionizing spectra may also be softer than standard AGN spectra \citep{wang_missing_2025}, and the timing of helium reionization limits how many LBDs can be highly super-Eddington \citep{cain_super-eddington_2026}. Our results support the conclusion of \citet{jiang_agns_2025} and \citet{mascia_ionizing_2026} that faint AGN are unlikely to dominate the ionizing photon budget during reionization.

\section{Summary}\label{sec:summary}

We measured Ly$\alpha$ in 12 CBEs at $z=4.17$--5.54 with DIVER JWST/NIRSpec spectroscopy, finding six detections and six upper limits. We learn three things.

\begin{enumerate}
    \item A visible BLR does not make Ly$\alpha$ bright: the Ly$\alpha$ strength does not follow H$\alpha$ or H$\alpha$/UV.

    \item The observed Ly$\alpha$ is unlikely to come directly from the BLR: every detected line is narrower than broad H$\alpha$.

    \item Ly$\alpha$ is more closely tied to the host galaxy than to the nucleus, with room for a nuclear contribution.
\end{enumerate}

The low Ly$\alpha$ visibility of these sources (median $f_{\rm esc,Ly\alpha}^{\rm eff}\approx0.09$) suggests modest ionizing escape, so faint AGN are unlikely to dominate reionization. Deep rest-UV spectroscopy and Ly$\alpha$ maps of larger samples will show how host and nuclear Ly$\alpha$ combine.

\begin{acknowledgements}
Y.Z.\ acknowledges support from the NIRCam Science Team contract to the University of Arizona, NAS5-02105. Y.Z.\ is also supported by JWST program \#6434. L.K.\ acknowledges the support of a Royal Society University Research Fellowship (grant number URF$\backslash$R1$\backslash$251793). A.J.B.\ acknowledges funding from the ``FirstGalaxies'' Advanced Grant from the European Research Council under the European Union's Horizon 2020 research and innovation programme (grant agreement No.\ 789056). Support for program \#6434 was provided by NASA through a grant from the Space Telescope Science Institute, which is operated by the Association of Universities for Research in Astronomy, Inc., under NASA contract NAS 5-03127. This manuscript benefited from language editing with ChatGPT (OpenAI, GPT-5.6 Sol), and the analysis code was developed with assistance from Claude Code (Anthropic, Claude Opus 5.5). All revised content and code were reviewed and verified by the authors.

This work is based on observations made with the NASA/ESA/CSA James Webb Space Telescope. The data were obtained from the Mikulski Archive for Space Telescopes at the Space Telescope Science Institute, which is operated by the Association of Universities for Research in Astronomy, Inc., under NASA contract NAS 5-03127 for JWST. These observations are associated with programs \#8018 (DIVER), \#1895 (FRESCO), \#3577 (CONGRESS), and \#1181 (JADES). JADES DR5 includes NIRCam data from JWST programs 1176, 1180, 1181, 1210, 1264, 1283, 1286, 1287, 1895, 1963, 2079, 2198, 2514, 2516, 2674, 3215, 3577, 3990, 4540, 4762, 5398, 5997, 6434, 6511, and 6541. JADES DR5 includes MIRI data from JWST programs 1180, 1181, and 1207. The authors acknowledge the teams of programs 1895, 1963, 2079, 2514, 3215, 3577, 3990, 6434, and 6541 for developing their observing program with a zero-exclusive-access period.

We respectfully acknowledge the University of Arizona is on the land and territories of Indigenous peoples. Today, Arizona is home to 22 federally recognized tribes, with Tucson being home to the O'odham and the Yaqui. The university strives to build sustainable relationships with sovereign Native Nations and Indigenous communities through education offerings, partnerships, and community service.
\end{acknowledgements}

\begin{contribution}
Y.Z.\ initiated and led this project, reduced the DIVER data, and led the interpretation. Z.M.\ performed the imaging decomposition and the host SED fitting. X.L.\ led the DIVER program. All authors contributed to the analysis and discussion.
\end{contribution}

\facilities{JWST (NIRSpec, NIRCam); MAST}

\software{
Astropy \citep{astropy_collaboration_astropy_2013, astropy_collaboration_astropy_2018, astropy_collaboration_astropy_2022},
JWST Calibration Pipeline \citep{bushouse_jwst_2025}
}

\end{document}